\documentclass[aps,prl,twocolumn,groupedaddress]{revtex4}
\usepackage{graphicx}
\begin{document}

\title{Quantum fluctuations in thin superconducting wires of finite length}

\author{H.P.\ B\"uchler, V.B.\ Geshkenbein, and G.\ Blatter}

\affiliation{Theoretische Physik, ETH-H\"onggerberg, CH-8093
Z\"urich, Switzerland}

\date{\today}

\begin{abstract}
In one dimensional wires, fluctuations destroy superconducting
long-range order and stiffness at finite temperatures; in an
infinite wire, quasi-long range order and stiffness survive at
zero temperature if the wire's dimensionless admittance $\mu$ is
large, $\mu > 2$. We analyze the disappearance of this
superconductor-insulator quantum phase transition in a finite wire
and its resurrection due to the wire's coupling to its environment
characterized through the dimensionless conductance $K$.
Integrating over phase slips, we determine the flow of couplings
and establish the $\mu$--$K$ phase diagram.
\end{abstract}

\maketitle

In systems with reduced dimensions (films and wires) fluctuations
and disorder strongly influence on the superconducting transition
temperature \cite{larkin99}, eventually driving a
superconductor--insulator (SI) transition which has attracted a
lot of attention recently \cite{haviland89,bezryadin00,lau01}. In
1D wires, the fluctuations of Cooper pairs appearing below the
mean-field transition at $T_{c0}$ define a finite resistance via
nucleation of thermally activated phase slips
\cite{langer67,newbower72} and hence remove the finite temperature
transition; superconductivity possibly survives only at zero
temperature. The focus then is on the quantum nucleation of phase
slips; their proliferation may trigger a zero temperature SI
quantum phase transition
\cite{giordano94,kashurnikov96,zaikin97,hekking97}. While first
attempts to observe quantum phase slips \cite{giordano94} are
still debated due to the granular structure of the wires
\cite{renn96,xiong97}, recent experiments on amorphous ultra-thin
wires \cite{bezryadin00,lau01} carry the signatures of a SI
transition in a homogeneous system. In this letter, we analyze the
bosonic Cooper pair fluctuations in realistic wires of finite
length which are naturally coupled to their environment through
their boundaries; we demonstrate how the SI transition is quenched
in the finite system and reappears through its coupling to the
environment.

\begin{figure}[htbp]
\includegraphics[scale=0.30]{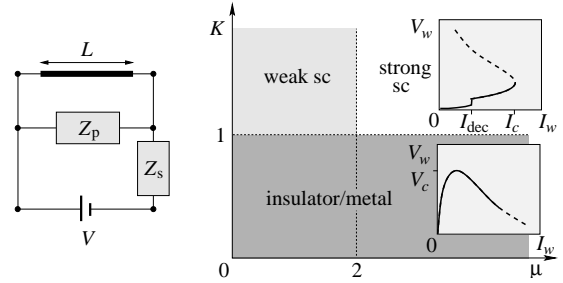}
\caption{Left: Setup with a quantum wire embedded in a voltage
   ($V$) driven loop with parallel ($Z_{\rm p}$) and serial
   ($Z_{\rm s}$) impedances defining the environment. Right: Phase
   diagram with superconducting and insulating phases
   separated by a quantum phase transition at $K=1$.
   The superconducting phase splits into weak and strong
   regimes separated by a crossover at $\mu \approx 2$,
   the leftover of the SI transition in the infinite wire.
   The insets show sketches of the wire's $I_{\rm w}$-$V_{\rm w}$
   characteristic at $T=0$. In the superconductor ($K > 1$)
   the algebraic characteristic is dominated by the environment
   at small currents $I<I_{\rm dec}$. A highly conducting shunt
   with $K\gg\mu$ allows to probe the wire above the deconfinement
   current $I_{\rm dec}$; $I_c$ denotes the critical current
   of the wire. The insulator exhibits a Coulomb gap behavior
   below the critical voltage $V_c$.}\label{setup}
\end{figure}

Previous studies of bosonic fluctuations in infinite wires
\cite{zaikin97} have found a $T=0$ SI quantum phase transition at
the critical value $\mu = 2$ of the dimensionless admittance
characterizing the wire's superconducting properties. At $T>0$ the
system exhibits a finite resistance; ignoring an additional
dissipative channel due to excited quasi-particles, we call the
non-superconducting state an insulating one. On the other hand,
real experiments are carried out on wires of finite length, $L
\sim 0.5-2~\mu$m typically \cite{bezryadin00}. The wire's coupling
to the environment through appropriate boundary conditions imposes
a drastic change in the phase slip dynamics and modifies the
wire's low-energy physics. A generic description is obtained by
embedding the wire in a voltage driven loop with impedances
$Z_{\rm p}(\omega)$ and $Z_{\rm s}(\omega)$ placed in parallel and
in series, see Fig.\ \ref{setup}. Here, we are mainly interested
in thermodynamic aspects (involving the static resistances $R_{\rm
p}$ and $R_{\rm s}$); furthermore we concentrate on the
current-driven limit with $R_{\rm s} \gg R_{\rm p}$, then
$I=V/R_{\rm s}$ is fixed. The voltage driven case with $R_{\rm p}
\ll R_{\rm s}$ and other mixed cases ($R_{\rm s}$ of order $R_{\rm
p}$) are easily derived from the current-driven solution via
Kirchhoff's laws, see below.

The coupling to the environment changes the $T=0$ phase diagram of
the infinite system, see Fig.\ \ref{setup}: The SI transition at
$\mu = 2$ is turned into a crossover, while a new quantum phase
transition appears at the critical value $K = R_{Q}/R_{\rm p}=1$
of the shunt's conductance ($R_{Q}=\pi \hbar/2 e^2$ denotes the
quantum resistance). A well conducting shunt with $K > 1$ relaxes
the strain on the wire and produces a superconducting response
with an algebraic $I$--$V$ characteristic, while a low conductance
$K < 1$ leads to the proliferation of phase slips and hence to an
insulator. The characteristic response of the superconducting wire
can be probed with a high conductance environment $K \gg \mu$ at
large drive above the current $I_{\rm dec}$ where confined
phase-slip pairs are separated.

In thin superconducting wires the plasma mode acquires a linear
dispersion \cite{mooij85,camarota01} with sound velocity $c_{s}$
(this contrasts with the bulk case where the Coulomb interaction
lifts the plasma mode to finite frequencies). The low frequency
action describing the bosonic fluctuations of the Cooper pairs
then takes the form \cite{kashurnikov96,zaikin97,hekking97}
\begin{equation}
   \frac{{\mathcal S}_{\rm w}}{\hbar} =
   \frac{\mu}{2\pi}
   \int_{0}^{\hbar \beta} \!\! d\tau\int_{-L/2}^{L/2}\! dx
   \left[c_{s} \left(\partial_{x} \phi \right)^{2}
   +\frac{1}{c_s}\left( \partial_{\tau} \phi\right)^{2} \right];
   \label{sact}
\end{equation}
the first (second) term accounts for the kinetic (compression)
energy (we assume a high energy cutoff $\hbar/\kappa$ limiting the
validity of (\ref{sact})). Within a mean-field description the
dimensionless admittance $\mu$ and the sound velocity $c_s$ are
related to the 1D superfluid density $\rho_s = n_s S/2 m^*$ and
the Coulomb interaction between the Cooper pairs
\cite{zaikin97,mooij85}, $\hbar \mu c_s/2\pi = \hbar^2 \rho_s/2$
and $\hbar\mu/2\pi c_s = (\hbar/2e)^2 C/2$ (here, $S$ is the
wire's cross section and $C = \epsilon/\ln(d^2/S)$ is the
capacitance per unit length, with $\epsilon$ the dielectric
constant of the surrounding media placed a distance $d$ away).
With $\lambda_{L}=(4\pi n_{s}e^{2}/m c^2)^{1/2}$ the London
penetration length we obtain $\mu \approx 30 \sqrt{C
S}/\lambda_{L}$ and $c_{s}^{2}/c^{2}\approx 0.1
S/\lambda_{L}^{2}C$. Going beyond the mean-field level, both
fermionic and bosonic high-energy fluctuations play a critical
role in the proper determination of the low-energy action
\cite{oreg}; here, we assume such effects to be included in our
choice of effective phenomenological parameters $\mu$ and $c_s$.

The coupling between the superconducting wire and the environment
involves the boundary fields $\phi_\pm (\tau)\equiv\phi_{\pm}
(x,\tau)|_{x=L/2}$ where $\phi_{\pm}(x,\tau) =\phi(x,\tau) \pm
\phi(-x,\tau)$: fluctuations in the phase difference $\phi_{-}$
generate a voltage $V_{\rm w}$ across the wire inducing currents
in the parallel shunts, while fluctuations in $\phi_{+}$ account
for charge accumulation. Here, we concentrate on the
current-driven setup (see Fig.~\ref{setup}) described by the
low-energy action
\begin{equation}
   \frac{{\mathcal S}_{\rm e}}{\hbar}
   =\frac{K}{2 \pi} \! \int
   \frac{d\omega}{4 \pi} |\omega|
   \left|\phi_{-}\left(\omega\right)\right|^{2}
   +\! \int_{0}^{\hbar\beta} \!\! d\tau
   \frac{I}{2 e} \phi_{-},
   \label{envact}
\end{equation}
with an ideal current source ($R_{\rm s} \gg R_{\rm p}$) driving
the system with the current $I$ and a parallel resistor with
resistance $R_{\rm p}$ accounting for the dissipation
($K=R_Q/R_{\rm p}$ is the dimensionless conductance of the shunt).
Protecting the system from the measurement setup using appropriate
filtering \cite{penttila99}, we can neglect capacitive effects.
%here, we ignore capacitive effects, see Ref.\ \cite{future} for a
%more detailed discussion).
The environment (\ref{envact}) does not account for an intrinsic,
e.g., quasi-particle induced, dissipation in the superconducting
wire; the latter contributes to the precise values of the
parameters $\mu$, $c_{s}$, and $\lambda$ (the vortex fugacity, see
below).% \cite{zaikin97}.

The statistical mechanics of the system `wire plus environment' is
determined by the partition function
\begin{equation}
  {\mathcal Z} = \int D[\phi]
  \exp\left[-{\mathcal S}(\phi)/\hbar\right]
  \label{partfunc}
\end{equation}
with ${\mathcal S} = {\mathcal S}_{\rm w} + {\mathcal S}_{\rm e}$.
Its main contributions arise from the combination of Gaussian and
topological fluctuations, so-called phase slips or instantons.
Gaussian fluctuations destroy the superconducting long-range order
as expressed by the logarithmically diverging phase correlator
$\langle \left[\phi (x,\tau) -\phi(x,\tau')\right]^2\rangle \sim
K^{-1} \ln|\tau-\tau'|$; however, the surviving quasi-long range
order is sufficient to allow for a finite phase stiffness, i.e.,
superconducting response. The latter is destroyed by the
proliferation of phase slips, the process we are going to analyze
in detail now.

Quantum phase slips are vortex-like solutions in $x,\tau$-space
with finite winding ($\nu = \pm 1$) around a core region of size
$x_{c} \leq c_{s} \kappa$ and $\tau_{c} \leq \kappa$ where the
superconducting gap $\Delta(x,\tau)$ drops to zero; outside this
core region they satisfy the differential equation $[c_{s}^{2}
\partial_{x}^{2} +\partial_{\tau}^{2}] \phi(\tau,x)=0$ within the
wire and respect the boundary conditions
\begin{eqnarray}
   K |\omega| \phi_{-} + \mu c_{s} [\partial_x \phi]_{-} =
   \pi I/e, \hspace{30pt}
   [\partial_x \phi]_{+}=0.
   \label{currentconservation}
\end{eqnarray}
The first equation describes current conservation as given by
Kirchhoff's law, with the supercurrent $I_{\rm w} = (e\mu
c_s/\pi)$ $[\partial_x \phi]_{-}$ through the wire and the
dissipative current $V_{\rm w}/R_{\rm p}$ in the shunt adding up
to the total external current $I$ (here, $[\partial_x
\phi]_\pm(\tau) \equiv
\partial_{x} \phi_\pm (x,\tau)|_{x=L/2}$ and $V_{\rm w} = \hbar |\omega|
\phi_{-}/2e$ is the voltage over the wire). The second equation
accounts for charge neutrality; we ignore additional capacitive
contributions here. Note that Kirchhoff's law maintains the form
(\ref{currentconservation}) when going over to the general
situation with impedances both in series and in parallel, $I_{\rm
w} + V_{\rm w}/R_{\rm t} = V/ R_{\rm s}$ with $R_{\rm
t}^{-1}=R_{\rm p}^{-1}+R_{\rm s}^{-1}$ the total conductance and
the total current $I = V/R_{\rm s}$ is expressed through the
driving voltage $V$. Hence once we know the solution $V_{\rm w}
(I) \equiv R_Q \chi_I(I)$ to the current-driven problem (with
$R_{\rm p} = R$), the solution $I_{\rm w}(V) \equiv \chi_V(V/R_Q)$
to the voltage-driven situation (with $R_{\rm s} = R$) follows
from the relation $I-K \chi_I (I)=\chi_V (I/K)$, with $K = R_Q/R$.

The partition function (\ref{partfunc}) factorizes into Gaussian
and topological parts, ${\mathcal Z}={\mathcal Z}_{\rm
\scriptscriptstyle G} {\mathcal Z}_{\rm  \scriptscriptstyle top}$,
where the latter can be expanded in a series ${\mathcal Z}_{\rm
\scriptscriptstyle top} = \sum_{n=0}^{\infty} (\lambda^n/ n!)^2
{\mathcal Z}_{n}(G)$ with
\[
   {\mathcal Z}_{n}
   \!= \!\int  \!\frac{ \prod_{m}^{2n} d\tau_{m} dx_{m}}
   {\left(c_{s} \kappa^{2}\right)^{2n}}
   \exp\!\biggl[\sum_{i \neq j}\nu_{i}
   G\left(x_{i},\tau_{i},x_{j},\tau_{j}\right)\nu_{j}\biggr]
\]
describing a (neutral) gas of $n$ vortex--anti-vortex pairs, each
contributing with a dimensionless action $G = S_{\rm pair}/\hbar$
(we assume $I=0$); in addition, each vortex-pair is weighted with
the fugacity $\lambda^2$ accounting for the microscopic structure
of the cores. Here, we assume that this factor is small, $\lambda
\sim \exp(-A R_{Q} L /R_{N}\xi) \ll 1$ with $R_{N}$ the normal
resistance of a wire, $\xi$ the superconducting coherence length,
and $A$ a factor of order unity \cite{zaikin97}.

The interaction $G$ between vortex--anti-vortex pairs is the
crucial quantity determining the system's behavior --- here we
only quote the relevant results, see Ref.\ \cite{future} for
details. For the infinite wire, the interaction between vortices
$\phi_k = \pm \arg[(x-x_k)+i\,c_s(\tau-\tau_k)]$, $k=1,2$, is
logarithmic at all distances $\overline{x}=x_{1}-x_{2}$,
$\overline{\tau} = \tau_{1} - \tau_{2}$,
\begin{equation}
   G(\overline{x},\overline{\tau})=
   \mu \ln [(\overline{x}^{2}/c_s^2
   +\overline{\tau}^{2})/\kappa^{2}].
   \label{infiniteaction}
\end{equation}
A standard Kosterlitz--Thouless (KT) scaling analysis provides the
RG equations \cite{kosterlitz74} $\partial_{l} \mu= -4 \pi^{2}
\mu^{2} \lambda^{2}$ and $\partial_{l} \lambda = (2-\mu) \lambda$
($l$ is the scaling parameter) and the system undergoes a
Berezinskii-Kosterlitz-Thouless transition \cite{kosterlitz74} at
$\mu = 2$. For $\mu > 2$, the number of free vortex--anti-vortex
pairs is quenched and the wire is in the superconducting phase;
for $\mu < 2$ the vortex fugacity increases and free
vortex--anti-vortex pairs proliferate
--- the system turns insulating.

\begin{figure}[hbtp]
   \includegraphics[scale=0.4]{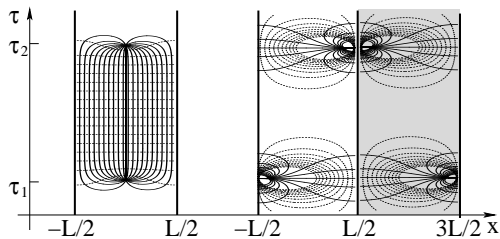}
   \caption{Phase slip solutions with periodic (solid lines
   denote contours $\phi_{\rm \scriptscriptstyle P}= {\rm const}.$,
   left) and Neumann ($\phi_{\rm \scriptscriptstyle N}$, right)
   boundary conditions at intervortex distance $L/c_s<\overline{\tau}$.
   Left: for $K \gg \mu$, defects
   are screened mutually, the $2\pi$ phase drop appears along the
   $x$-axis and drives a large current (dashed) through the highly
   conducting shunt; the resulting string confines defect pairs.
   Right: with $\mu \gg K$, defects are screened individually by their
   mirror images (grey underlaid), the $2\pi$ phase drop appears along
   the $\tau$-axis and sets up displacement currents in the
   wire which cannot escape through shunt, hence charge accumulates
   at the boundary resulting in a large voltage over the shunt;
   the defects are asymptotically free.
   \label{neumanBC2}}
\end{figure}

In a finite wire the solution has to respect the boundary
condition (\ref{currentconservation}). At low frequencies and
vanishing drive these reduce to the Neumann boundary conditions,
$\partial_{x}\phi_{{\rm \scriptscriptstyle N}}(\pm L/2,\tau) = 0$,
i.e., the currents induced by phase slips cannot leave the wire.
The solution $\phi_{{\rm \scriptscriptstyle N}}(x,\tau)$ derives
from the $2L$-periodic solution $\phi_{\scriptscriptstyle 2L-{\rm
p}}$ using mirror vortices, $\phi_{{\rm\scriptscriptstyle
N}}(x,\tau)=\phi_{\scriptscriptstyle 2L-{\rm p}}(x,\tau) +
\phi_{\scriptscriptstyle 2L-{\rm p}}(L-x,\tau)$; in turn, the
instanton solution with periodic boundary conditions
$\phi_{\scriptscriptstyle 2L-{\rm p}}(x,\tau) =
\phi_{\scriptscriptstyle 2L-{\rm p}}(x+2L,\tau)$ derives from the
solution in the infinite wire via conformal transformation. Fig.\
\ref{neumanBC2} illustrates the geometry of the phase slips; most
remarkably, at large distances the mutual screening of the
defect-pair is replaced by the screening via image charges. Hence
the wire's contribution $G_{\rm w}$ to the action saturates at
inter-vortex separations $\overline{\tau} \equiv |\tau_2-\tau_1| >
L/\pi c_{s}$, resulting in asymptotically free vortices.
Furthermore, their action can be minimized by moving them to the
boundary such that $G_{\rm w} = 0$.

In this situation the pair interaction is determined by the
environment alone. Inserting the boundary field $\phi_{{\rm
\scriptscriptstyle N}-}(\tau)$ describing two opposite kinks of
shape $\pm 2\arctan\{\sinh[(\pi c_{s}/L)(\tau\!-\!\tau_{k})]
/\cos(\pi x_{k}/L)\}$, $k=1,2$, into the action (\ref{envact}) we
obtain the contribution
\begin{equation}
   G_{\rm e}(x_{1},x_{2},\overline{\tau})
   \approx K \ln\biggl[\frac{\cos^{\, 2}\!
   \frac{\pi}{2L}(x_{1} +x_{2})
   +\left( \frac{\pi c_s}{2L} \overline{\tau}\right)^{2}}
   {\cos \frac{\pi}{L} x_{1} \> \cos \frac{\pi}{L} x_{2}}\biggr]
   \label{GE}
\end{equation}
which diverges logarithmically at long time scales, but with a
weight determined by the dimensionless conductance $K$ of the
parallel shunt. In the end, the largest contribution to the
partition function arises from vortices nucleated at the boundary
with an interaction $G \approx G_{\rm e} = K \ln
(\overline{\tau}^{2}/\kappa^{2})$ and the problem maps to a system
of charged particles in 1D with fugacity $\lambda$. The
corresponding RG equations \cite{bulgadaev84} take the form
$\partial_{l} K = 0$ and $\partial_{l}\lambda = (1-K) \lambda$; in
contrast to the infinite wire, the prefactor $K$ of the logarithm
is invariant under the RG flow and we obtain a transition at $K=1$
\cite{schmid83}, with a superconducting phase for $K>1$, while for
$K<1$ the vortices nucleated at the boundary drive the system
insulating.

In the analysis above we have determined the shape of the
vortex--anti-vortex pair assuming Neumann boundary conditions,
i.e., no currents can leave the wire. Inserting this approximate
solution back into (\ref{envact}) then has provided us with the
pair-interaction $G \approx G_{\rm e} =
2K\ln(\overline{\tau}/\kappa)$. In order to check the consistency
of this approximation we determine the correction $\delta
{\mathcal S}$ due to the finite current $K|\omega|\phi_-$ flowing
through the shunt (cf.\ (\ref{currentconservation})),
\begin{equation}
   \frac{\delta {\mathcal S}}{\hbar}
   = -\frac{K}{4 \pi} \int \frac{d\omega}{2 \pi}
   \frac{|\omega|\left|\phi_{{\rm\scriptscriptstyle N}-}\right|^2}
   {1+(\mu/K) \left| \coth(\omega L/2 c_{s})\right|},
   \label{correction}
\end{equation}
with $\phi_{{\rm\scriptscriptstyle N}-} \approx (4\pi/\omega)
\sin(\omega\overline{\tau}/2)$; performing the integral in
(\ref{correction}) shows that the corrections are indeed small at
low frequencies. For a setup with a poorly conducting shunt ($K
\ll \mu$) the corrections remain small at higher energies when
$\overline{\tau} < L/\pi c_{s}$, $\delta S \sim -\mu (K/\mu)^{2}
\ln(\overline{\tau}/\kappa)$.

However, in the opposite case $K \gg \mu$ where a highly
conducting shunt protects the superconductor the corrections are
large: in the intermediate regime $L/\pi c_s < \overline{\tau} <
\tau_{\rm \scriptscriptstyle B} \equiv K L/\pi \mu c_s$, we find
$\delta S \sim -2K \ln(\overline{\tau}/\kappa) + 2 \pi \mu c_s
\overline{\tau}/L$. The first term cancels the logarithmic
interaction with the environment, while the second term describes
linearly confined pairs. It is then appropriate to change
strategy: for $K \gg \mu$ and high frequencies the boundary
condition (\ref{currentconservation}) reduces to $\phi_{-}=0$ and
$[\partial_x\phi]_{+}=0$ for $I=0$ and the phase-field for the
pair is given by the $L$-periodic solution
$\phi_{\scriptscriptstyle L-{\rm p}}$, producing an interaction
between the vortices
\[
   G_{\rm \scriptscriptstyle p}(x,\tau)
   = \mu \ln\!
   \biggl[\!\biggl(\frac{L}{\pi c_{s}\kappa}\biggr)^{2}
   \biggl(\sinh^{2}\frac{\pi c_{s}\overline{\tau}}{L}
   +\sin^{2}\frac{\pi\overline{x}}{L}\biggr)\biggr]
\]
describing defect pairs linearly confined along the $\tau$-axis
for distances $\overline{\tau} > L/\pi c_s$, cf.\ Fig.\
\ref{neumanBC2}. At the same time, no voltage appears over the
shunt and the contribution from the environment vanishes, thus $G
\approx G_{\rm \scriptscriptstyle p}$. Hence, for a highly
conducting shunt $K \gg \mu$ the interaction starts with a
logarithm $G(\overline{\tau} < L/\pi c_s) \sim \mu \ln
[(\overline{\tau}^2+\overline{x}^2)/\kappa^2]$, goes over into
confinement $G(L/\pi c_s <\overline{\tau} < \tau{{\rm
\scriptscriptstyle B}}) \sim 2 \pi \mu c_s \overline{\tau}/L$, and
ends with the low-frequency behavior $G(\tau{{\rm
\scriptscriptstyle B}}< \overline{\tau}) \sim K\ln
(\overline{\tau}^2/\kappa^2)$ determined by the environment.

In the end, we find that for all values of $K$ the low-energy
physics of the system is determined by the environment. On the
other hand, the renormalization down to low energies depends on
the ratio $\mu/K$: for $K \ll \mu$ the interaction is entirely
determined by the environment, while for $K \gg \mu$ the
phase-slip pairs go through an intermediate regime of linear
confinement inducing an exponential drop of the phase-slip
fugacity, $\lambda(\tau_{\rm \scriptscriptstyle B}) \sim
\lambda(L/\pi c_{s}) \,\exp[-G_{\rm \scriptscriptstyle p}
(\tau_{\rm \scriptscriptstyle B})]=\lambda(L/\pi
c_{s})\,\exp(-K)$; this drop in $\lambda$ manifests itself in the
$I_{\rm w}$--$V_{\rm w}$-characteristic.

The above analysis results in a $T=0$ phase diagram as sketched in
Fig.\ \ref{setup}: The transition at $K=1$ separates a superfluid
phase for $K > 1$ from an insulating one at $K < 1$. In addition,
we distinguish between two different superconducting regimes at
$K>1$: for $\mu < 2$ the fugacity for the nucleation of small
($\overline{\tau}< L/\pi c_{s}$) vortex--anti-vortex pairs is
strongly increased as compared to the regime $\mu > 2$, a leftover
from the SI transition in the infinite wire which is relevant at
temperatures $T>\hbar \pi c_{s}/L$ and drives $I > 2 e \mu
c_{s}/L$ probing these small scales.

With the interaction between defect pairs turning logarithmic on
large scales $\overline{\tau} > \tau_{\rm \scriptscriptstyle
K}\equiv \max(\kappa,\tau_{\rm \scriptscriptstyle B})$, we find
that the low-energy physics of the wire reduces to that of a
Josephson junction with a parallel shunt $R$; the partition
function is equivalent to that of a particle in a periodic
potential with damping $\eta= K/2\pi$ \cite{schmid83}. The
$I$--$V$-characteristic of such Josephson junctions has been
studied in detail \cite{schmid83,schoen90}; below, we review the
main results. The restricted validity of the effective action
${\mathcal S} = {\mathcal S}_{\rm w} +{\mathcal S}_{\rm e}$ to low
energy scales limits this analysis to low temperatures $T <
\hbar/\tau_{\rm \scriptscriptstyle K}$ and low currents $I< 2e
K/\pi\tau_{\rm \scriptscriptstyle K}$.

Within the superconducting phase $K>1$, the $I-V$ characteristic
at low drives $I$ is calculated perturbatively in the vortex
fugacity $\lambda$ (with $\lambda(\tau_{\rm \scriptscriptstyle K})
\rightarrow \lambda$ the renormalized fugacity at $\tau_{\rm
\scriptscriptstyle K}$). At finite temperatures the two energy
scales $T$ and $\hbar\pi I/2eK$ define two regimes in the
response, $V_{\rm ps}(\pi I/2eK \ll T/\hbar) \sim \lambda^2 R_Q I
[T \tau_{\rm \scriptscriptstyle K}/\hbar]^{2K-2}$ and $V_{\rm ps}
(T/\hbar \ll \pi I/2eK) \sim (\hbar\lambda^2 /e \tau_{\rm
\scriptscriptstyle K}) [I\tau_{\rm \scriptscriptstyle
K}/e]^{2K-1}$; the algebraic characteristic at $T=0$ turns into a
linear response at finite $T$ and small drives $I$. We conclude,
that in 1D wires superconductivity survives only at $T=0$ and
under the condition of a good protection by a high conductance
shunt with $K > 1$. In addition, if $K\gg\mu$, the linear
confinement in the regime $c_s/L<\overline{\tau}<\tau_{\rm
\scriptscriptstyle B}$ produces a sharp voltage step $V_{\rm dec}
\sim (\hbar\lambda^{2}/e \kappa)(L/c_{s} \kappa) [\pi \kappa
c_{s}/L]^{2\mu-2}$ at the deconfinement current $I_{\rm dec} = 2 e
\mu c_{s}/L$. At high drives $I> I_{\rm dec}$ (and at temperatures
$T > \hbar \pi c_{s}/L$) we start probing the wire (i.e.,
intervortex distances $\overline{\tau} < \pi c_{s}/L$) and the
$I$-$V_{\rm w}$ characteristic is given by the result for the
infinite system \cite{zaikin97}, $V_{\rm ps}(\pi c_{s}/L < \pi I/2
\mu e \ll T/\hbar) \sim (\lambda^2 R_Q I) (L/c_s \kappa)
[T\kappa/\hbar]^{2\mu-3}$ and $V_{\rm ps}(\pi c_{s}/L < T/\hbar
\ll \pi I/2 \mu e) \sim (\hbar\lambda^2/e\kappa)
(L/c_s\kappa)[I\kappa/e\mu]^{2\mu-2}$. The wire's $I_{\rm
w}-V_{\rm w}$-characteristic derives from solution of the implicit
equation $V_{\rm w}(I_{\rm w}) = V_{\rm ps}(I=I_{\rm w}+V_{\rm
w}/R_{\rm p})$.

In the insulating phase $K < 1$ the phase slips describe tunneling
of the phase difference $\phi_{-}$ to neighboring states $\phi_{-}
\pm 2\pi$. In the limit $K \rightarrow 0$ this tunneling between
periodic minima is coherent and leads to the formation of a Bloch
band $\varepsilon(q)$ of width $W_{0} = \hbar \lambda/\kappa$,
with $q$ the quasi-momentum associated to $\phi_-$. Applying a
small current $I$, a voltage $2eV_{\rm w}=\hbar\langle
\dot{\phi}_- \rangle =\partial_q \varepsilon(q)$ is set up across
the wire and all the current flows over the parallel shunt
\cite{schoen90}; the system exhibits a linear response $V_{\rm w}
= R I$. This behavior remains valid for finite $K < 1$, but with a
renormalized band width \cite{schoen90,ingold99,penttila99} $W
\approx W_{0}[W_{0} \tau_{\rm \scriptscriptstyle
K}/\hbar]^{K/(1-K)}$, defining a linear response regime for small
voltages $2eV_{\rm w} < 2e V_c = \max(\partial_q \varepsilon) \sim
W$. At higher drives, the voltage $V_{\rm w}$ decreases as the
phase slips no longer block the wire and the current flows across
both channels.

% We note that the low-drive response of the
%superconducting ($K>1$) and insulating phases ($K<1$) are related
%via duality \cite{fendley}, $I-K\chi_I(I,K) = K\chi_I(\alpha
%I,1/K)$, with $\alpha$ a suitable scaling factor.

The SI quantum phase transition in the finite wire is different
from the one in the infinite system; it is of the type inherent to
finite systems coupled to a dissipative environment, e.g., the
resistively shunted Josephson junction \cite{schmid83} or the
dissipative two-state system \cite{leggett87}. The environment
determines the wire's thermodynamic state and hence the overall
shape of the $I_{\rm w}$--$V_{\rm w}$-characteristic, see Fig.\
\ref{setup}; the wire's parameters $\mu$ and $\lambda$ make their
appearance (through the voltage jump at $I_{\rm dec}$ and the
wire-dominated characteristic at large currents $I > I_{\rm dec}$)
only in the superconducting phase $K>1$ and for the case $K\gg\mu$
where a highly conducting shunt is protecting the wire. The
insulating phase is established at small values $K < 1$;
phase-slips drive the current over the parallel shunt, resulting
in an ohmic characteristic, until leakage through the wire sets in
at $V_c$. Here, we have assumed a small vortex fugacity $\lambda$
and hence the observation (and consistent interpretation) of the
characteristic voltage-peak requires a sensitive probe as $V_c
\propto \lambda^{1/(1-K)}$.

% \bibliographystyle{/home/buechler/Refbib/apsrev}
 %\bibliography{/home/buechler/Refbib/journals,/home/buechler/Refbib/ref}

%\end{multicols}
\end{document}